\documentclass[a4paper,12pt]{article}
\usepackage{amssymb,amsmath}
\usepackage{bm}%
\usepackage{stmaryrd}
\usepackage{slashed}
\usepackage{arydshln}
\usepackage[pdftex]{hyperref}
\hypersetup{
	colorlinks=true,
	citecolor=blue,
	linkcolor=blue,
	urlcolor=orange,
}
\usepackage{cite}
\usepackage{comment}

\def\Z2{\mathbb{Z}_2^2}
\def\PB#1#2{ \{#1,#2\}_{\mathrm{PB}} }
\def\DB#1#2{ \{#1,#2\}_{\mathrm{DB}} }
\def\nn{\nonumber}
\def\ph#1#2{(-1)^{\vec{#1}\cdot\vec{#2}}}
\def\g{\mathfrak{g}}

\begin{document}

\title{$\mathcal{N}=2$ Double graded supersymmetric quantum mechanics via dimensional reduction}

\author{N. Aizawa, Ren Ito, Toshiya Tanaka
	\\[10pt]
	Department of Physics, Graduate School of Science, \\ Osaka Metropolitan University, \\
	Nakamozu Campus, Sakai, Osaka 599-8531, Japan}

\maketitle
\thispagestyle{empty}

\vfill
\begin{abstract}
We present a novel $\mathcal{N} = 2 $ $\mathbb{Z}_2^2$-graded supersymmetric quantum mechanics ($\Z2$-SQM) which has different features from those introduced so far. 
It is a two-dimensional (two-particle) system and is the first example of the quantum mechanical realization of an eight-dimensional irrep of the $\mathcal{N}=2$ $\mathbb{Z}_2^2$-supersymmetry algebra. 
The $\Z2$-SQM is obtained by quantizing the one-dimensional classical system derived by dimensional reduction from the two-dimensional  $\Z2$-supersymmetric Lagrangian of $\mathcal{N}=1$, which we constructed in our previous work. The ground states of the $\Z2$-SQM are also investigated.
\end{abstract}

\clearpage
\setcounter{page}{1}

\section{Introduction}

In our previous work \cite{AiItoTa}, a $\Z2$-graded supersymmetric Lagrangian in two-dimensional spacetime was constructed by the $\Z2$-extension of the superfield formalism. 
The $\Z2$-supersymmetry is a higher graded extension of the supersymmetry, based on the $\Z2$-graded superalgebras introduced by Bruce \cite{Bruce}. 
The Lagrangian given in \cite{AiItoTa}, which is $\mathcal{L}$ in \eqref{Lint2D} of the present paper, has a very general interaction terms and appropriate choices of them give $\Z2$-graded supersymmetric extensions of the two-dimensional integrable systems, e.g., sine(h)-Gordon equation and Liouville equation.  
The integrability of these $\Z2$-extended equations is an open problem, however one may expect the existence of a novel class of integrable systems characterized by the $\Z2$-supersymmetry. 
Indeed, a $\Z2$-graded extension of the sine-Gordon equation, which is different from the one in \cite{AiItoTa}, was introduced and its integrability is shown by Bruce \cite{bruSG} and this is the only integrable classical systems having the $\Z2$-supersymmetry known so far. 
Therefore, in order to open up a new field of integrable systems, the study of the classical systems obtained form $\mathcal{L}$ in \eqref{Lint2D} is important. 

It is also important to quantize the Lagrangian $\mathcal{L}$ which will gives quantum integrable systems.  
In the present paper, however, instead of quantizing $\mathcal{L},$ we study the simpler but highly non-trivial case, that is, $\Z2$-supersymmetric quantum mechanics ($\Z2$-SQM) obtained from $\mathcal{L} $ via dimensional reduction. 
The $\Z2$-SQM was first introduce by Bruce and Duplij \cite{BruDup} which is $\mathcal{N} = 1 $ in our terminology (see \S \ref{SEC:Prel} for the definition of $\mathcal{N}$). 
The operators of this $\Z2$-SQM close in ordinary 1D super-Poincar\'e algebra as well as its $\Z2$-counterpart. However, this does not means the $\Z2$-SQM is trivial, since the $\Z2$-SQM shows detectable difference from the ordinary SQM in multiparticle sectors \cite{Topp,Topp2}. 
The Bruce-Duplij $\Z2$-SQM is also extended to $ \mathcal{N} > 1 $ \cite{AAD}, $\mathbb{Z}_2^n$-grading \cite{AAd2} and conformal symmetries \cite{DoiAi1}. 

Our Lagrangian $\mathcal{L}$ in \eqref{Lint2D} is $\mathcal{N} = 1 $ and defined in two-dimensional spacetime. Reduction of it to one-dimension gives $\mathcal{N} = 2 $ $\Z2$-supersymmetric classical mechanics. 
We quantize the 1D system using a $\Z2$-graded extension of the Dirac-Bergmann method of constrained systems \cite{AKTcl,AKTqu}.  
This gives us a $\Z2$-SQM which has  different features from the $\Z2$-SQMs mentioned above  (see \S \ref{SEC:QM}).  
In particular, the $\Z2$-SQM obtained is a two-dimensional or two-particle (with the same mass) quantum mechanics and the left and right movers are separated in the light cone coordinates. 
Furthermore, it is realized by $ 8 \times 8 $ matrix differential operators which corresponds to the eight dimensional irrep of $\mathcal{N} = 2 $ $\Z2$-supersymmetry algebra \cite{AiDoi} and it is the first example of the quantum mechanical realization of the irrep.

It is well known that SQMs are closely related to solvable potentials through the factorization of Hamiltonian (see e.g., \cite{Junker,Bagchi}). 
It is also known that $\Z2$-graded algebraic structure appears in simple solvalbe systems in quantum mechanics \cite{BeDe,AKTT1,AKTT2}. 
We therefore expect the $\Z2$-SQMs to have a deep connection with solvable quantum mechanical systems.

Before proceeding further, we mention some works discussing $\Z2$-graded algebraic structure in physics. Vasiliev pointed out that the symmetry group of SUGRA in de Sitter spacetime is enhanced to $\Z2$-graded superalgebra \cite{Vas}. 
The quasi-spin formalism  is generalized to higher graded algebra in \cite{jyw} and superconformal symmetry in two-dimension is also generalized to $\Z2$-graded setting \cite{zhe}. 
Equivalence between algebraic structures generated by parastatistics triple relations of Green and Greenberg–Messiah, and certain orthosymplectic $\Z2$-graded superalgebras is pointed out in  \cite{tol}. This observation of $\Z2$-graded superalgebras in parastatistics leads the  further development of parastatistics representations of $\Z2$-graded superalgebras \cite{StoVDJ,StoVDJ2,StoVDJ3}.  
We also comment that the paraparticles are simulated recently by using a trapped ion \cite{ParaP}.

There are some proposals of $\mathbb{Z}_2^n$-graded extensions of the spacetime supersymmetry \cite{LR,Toro1,Toro2,tol2} which are related to higher graded SQMs. 
Regarding the higher graded supersymmetry, we mention the bosonization \cite{Quesne}, sigma model \cite{brusigma} and $n$-bit extension of parastatistics \cite{BdFRT}.  
A precise analysis of the $\Z2$-graded superfield formulation of $\Z2$-supersymmetry has recently been done  \cite{DoiAi2,AiDoi,AIKT}. 
The $\Z2$-graded superfield formulation is a simplest example of higher supergeometry which was started in \cite{CGP1} (see \cite{Pz2nint,PonSch} for a concise review of the higher supergeometry). 
Integration over the $\Z2$-superspace is a necessary ingredient of the superfield formulation. 
There are some different ideas of integration and one of them recently proposed  by two of the present authors is suitable for the superfield formulation \cite{NARI}. 

This paper is organized as follows: 
In the next section, we recall the definition of the $\Z2$-graded Lie superalgebras and collect the results from \cite{AiItoTa} which we need in the present work. 
In \S \ref{SEC:cl}, we investigate the classical aspects of the 1D system obtained by dimensional reduction. 
The 1D Lagrangian is derived from the 2D one and equations of motion and conserved Noether charges are computed explicitly. 
The Lagrangian is singular but all the constraints are second class. We thus develop a $\Z2$-extension of the Dirac-Bargmann method suitable to the present model to quantize the system. We also observe the increase of the  $\Z2$-supersymmetry from $\mathcal{N} = 1 $ to $ \mathcal{N} = 2. $ 
\S \ref{SEC:QM} is devoted to the study of the quantized system. The quantum operators are realized in terms of the eight-dimensional real irrep of the Clifford algebra $Cl(4,2)$. The use of light cone coordinates provides separation of variables. This allows us to easily study the ground states of the Hamiltonian. 
We close the paper with a short summary and some remarks in \S \ref{SEC:CR}

\setcounter{equation}{0}
\section{Preliminaries} \label{SEC:Prel}

Let us first recall the definition of $\Z2$-graded Lie superalgebras \cite{RW1,RW2} (see also \cite{Ree,sch1}). 
A $\Z2$-graded vector space  (over $\mathbb{R}$ or $ \mathbb{C}$) is the direct sum of homogeneous vector subspaces labeled by an element of $\Z2$:
\[
\g = \g_{(0,0)} \oplus \g_{(1,1)} \oplus \g_{(1,0)} \oplus \g_{(0,1)}.
\]
An element of $ \g_{\vec{a}}$  is said to have the $\Z2$-\textit{degree} $ \vec{a}  \in \Z2.$ 
We define the $\Z2$-Lie bracket by
\begin{equation}
	\llbracket X, Y \rrbracket = XY - (-1)^{\vec{a}\cdot\vec{b}} Y X,
	\quad
	X \in \g_{\vec{a}}, \ Y \in \g_{\vec{b}}
\end{equation}
where $ \vec{a}\cdot\vec{b} $ is the standard scalar product of two dimensional vectors. 
Namely, the $\Z2$-Lie bracket is the commutator (anti-commutator) for $ \vec{a}\cdot\vec{b} $ is even (odd). 
A $\Z2$-graded vector space is said to be a $\Z2$-graded Lie superalgebra if 
$ \llbracket X, Y \rrbracket \in \g_{\vec{a}+\vec{b}} $ and the Jacobi identity is satisfied:
\[
\llbracket X, \llbracket Y, Z \rrbracket \rrbracket
= \llbracket \llbracket X, Y \rrbracket, Z \rrbracket
+ (-1)^{\vec{a}\cdot \vec{b}} \llbracket Y, \llbracket X, Z \rrbracket \rrbracket.
\]
If $ \llbracket X, Y \rrbracket = 0, $ we say that $ X $ and $Y$ are $\Z2$-\textit{commutative}. 
We also define the even and odd subspaces of $\g$ by $ \g_{(0,0)} \oplus \g_{(1,1)} $ and $\g_{(1,0)} \oplus \g_{(0,1)},$ respectively.  

The $\Z2$-graded Lie superalgebra considered in \cite{AiItoTa}, which is denoted simply by $\g$, is five-dimensional and the $\Z2$-degree assignment is as follows:
\begin{equation}
	H \in \g_{(0,0)}, \quad Z, \ L_{11} \in \g_{(1,1)}, \quad Q_{10} \in \g_{(1,0)}, \quad Q_{01} \in \g_{(0,1)}
\end{equation}
Their non-vanishing $\Z2$-Lie brackets  in terms of commutator or anticommutator are given by
\begin{alignat}{2}
	\{Q_{10},Q_{10}\}&=\{Q_{01},Q_{01}\}=2H,
     &
    [Q_{10},Q_{01}] &=iZ,
    \nn\\
    [L_{11}, H] &= \frac{i}{2}Z, & [L_{11}, Z] &= 2iH,
	\nonumber \\
	\{L_{11}, Q_{10}\} &=-\frac{1}{2}Q_{01}, &\quad \{ L_{11}, Q_{01}\} &= \frac{1}{2}Q_{10}.
	\label{SUSY-Lorentz-Alg}
\end{alignat}
The subalgebra $ \langle \ H, Q_{10}, Q_{01}, Z \ \rangle $ is the $\Z2$-supersymmetry algebra introduced in \cite{Bruce}. 
We refer this algebra as $ \mathcal{N} = 1 $ since each odd subspace has only one element. 

We consider the eight real fields with $\Z2$-grading defined in two-dimensional spacetime
\begin{align}
	& \varphi_{00}(t,x),\quad A_{00}(t,x),\quad  A_{11}(t,x),\quad \varphi_{11}(t,x),
	\nn \\
	& \psi_{10}(t,x),\quad \lambda_{10}(t,x),\quad \psi_{01}(t,x),\quad \lambda_{01}(t,x)
\end{align}
where the suffices indicate their $\Z2$-degree and the fields are $\Z2$-commutative. 
It is shown in \cite{AiItoTa} that the following action is invariant under the transformations generated by $\g:$
\begin{align}
	S &= \int dt\, dx\, \mathcal{L}, \qquad \mathcal{L}= \mathcal{L}_{kin} + \mathcal{L}_{int},
	\nn \\
	\mathcal{L}_{kin} &=  \frac{1}{2} (\dot{\varphi}^2_{00}  - \varphi_{00}'{}^2 + \dot{\varphi}^2_{11}  - \varphi_{11}'{}^2) 
	+2A_{00}^2 + 2A_{11}^2
	\nonumber \\
	&
	+i (\psi_{10} \dot{\psi}_{10} + \psi_{01} \dot{\psi}_{01} + \lambda_{10} \dot{\lambda}_{10} + \lambda_{01} \dot{\lambda}_{01})
	\nonumber \\
	& -i(\psi_{10} \lambda_{10}' - \psi_{10}' \lambda_{10} -\psi_{01} \lambda_{01}' + \psi_{01}' \lambda_{01}),
	\nonumber \\
	\mathcal{L}_{int} &= -2\alpha \big( A_{11} V_{00} + A_{00} V_{11}\big) 
	\nn \\
	&+2\alpha \big(  (\psi_{10} \psi_{01} + \lambda_{10} \lambda_{01} ) \partial_{00}V_{00}
	+ i (\psi_{10} \lambda_{10} + \psi_{01}\lambda_{01}) \partial_{00}V_{11} \big)
	\label{Lint2D}
\end{align}
where $ \alpha $ is a degree $(1,1)$ coupling constant and $ V_{00}, V_{11}$ are functions of $ \varphi_{00}, \varphi_{11}$ satisfying
\begin{equation}
	\partial_{00} V_{00}(\varphi_{00},\varphi_{11}) = \partial_{11} V_{11}(\varphi_{00},\varphi_{11}), \quad 
	\partial_{11} V_{00}(\varphi_{00},\varphi_{11}) = \partial_{00} V_{11}(\varphi_{00},\varphi_{11}) 
	\label{ConstraintV}
\end{equation}
with 
\begin{equation}
	\partial_{00} := \frac{\partial}{\partial \varphi_{00}}, \qquad \partial_{11} := \frac{\partial}{\partial \varphi_{11}}.
\end{equation}
$ H$ and $Z$ are the generator of the translations of $t$ and $ x$, respectively. 
$ Q_{10} $ and $ Q_{01}$ are supercharges mixing up even (bosonic) and odd (fermionic) fields and changing the degree by $ (1,0) $ and $(0,1)$, respectively. 
$ L_{11}$ is the degree $(1,1)$ Lorentz transformation which gives rise to mixture among bosonic (fermionic) fields with different degrees. 
The transformation generated by $ Z$ and $ L_{11} $ disappear after the reduction to one-dimensional spacetime due to the non-existence of space translationa and Lorentz transformation.  
Explicit form of the transformations are given in the equations (3.28)--(3.32) of \cite{AiItoTa}. 
The matrix presentation of the generators is found in (3.34)--(3.37) of \cite{AiItoTa}. 

As is seen from the Lagrangian \eqref{Lint2D}, $ A_{00}, A_{11}$ are auxiliary, i.e., their equations of motion are given by the algebraic equation
\begin{equation}
	A_{00} = \frac{\alpha}{2}V_{11}, \qquad A_{11} = \frac{\alpha}{2}V_{00}. \label{EoMA}
\end{equation}
Using these relations, we remove the auxiliary fields.

\section{One-dimensional model : classical mechanics} \label{SEC:cl}
\setcounter{equation}{0}

\subsection{Lagrangian and equations of motion}

We make the dimensional reduction $ (t,x) \to (t)$. 
Then, we have the worldline $\Z2$-supersymmetric Lagrangian from \eqref{Lint2D} 
\begin{align}
    \mathcal{L} &=  \frac{1}{2} (\dot{\varphi}^2_{00}  + \dot{\varphi}^2_{11} ) 
	+i (\psi_{10} \dot{\psi}_{10} + \psi_{01} \dot{\psi}_{01} + \lambda_{10} \dot{\lambda}_{10} + \lambda_{01} \dot{\lambda}_{01})	
	\nonumber \\
	&
	+2A_{00}^2 + 2A_{11}^2
	 -2\alpha \big( A_{11} V_{00} + A_{00} V_{11}\big) 
	\nn \\
	&+2\alpha \big(  (\psi_{10} \psi_{01} + \lambda_{10} \lambda_{01} ) \partial_{00}V_{00}
	+ i (\psi_{10} \lambda_{10} + \psi_{01}\lambda_{01}) \partial_{00}V_{11} \big).
\end{align}
This Lagrangian is invariant under the following transformations generated by $\g$ which is the one-dimensional reduction of (3.28)--(3.32) of \cite{AiItoTa}:

\medskip\noindent
(i) transformations by $H$ and $Z$
\begin{equation}
	\delta_{00} f(t,x) = -\frac{\epsilon_{00}}{2}\partial_t f(t,x), 
	\quad 
	\delta_{11} f(t,x) = 0, \quad
	\qquad \text{for any component fields} \label{NewVariTrans0}
\end{equation}
(ii) transformation by $Q_{10}$
\begin{align}
	\delta_{10} \varphi_{00}=&-i\epsilon_{10}\psi_{10}
	,&
	\delta_{10} \varphi_{11}=&\epsilon_{10}\lambda_{01}
	,\nonumber \\
	\delta_{10} \psi_{10}=& \frac{1}{2}\epsilon_{10} \dot\varphi_{00}
	,&
	\delta_{10} \lambda_{01}=&-\frac{i}{2}\epsilon_{10}\dot{\varphi}_{11}
	,\nonumber \\
	\delta_{10} \psi_{01}=&i\epsilon_{10} A_{11}
	,&
	\delta_{10} \lambda_{10}=&\epsilon_{10}
	A_{00}
	,\nonumber \\
	\delta_{10}  A_{11} =&-\frac{1}{2}\epsilon_{10}
	\dot\psi_{01}
	,&
	\delta_{10}  A_{00} =&-\frac{i}{2}\epsilon_{10}
	\dot{\lambda}_{10},  \label{NewVariTrans2}
\end{align}
(iii) transformation by $Q_{01}$
\begin{align}
	\delta_{01} \varphi_{00}=&-i\epsilon_{01}\psi_{01}
	,&
	\delta_{01} \varphi_{11}=&\epsilon_{01}\lambda_{10}
	,\nonumber \\
	\delta_{01} \psi_{10}=&i\epsilon_{01} 
	A_{11}
	,&
	\delta_{01} \lambda_{01}=&\epsilon_{01}
	A_{00}
	,\nonumber \\
	\delta_{01} \psi_{01}=& \frac{1}{2}\epsilon_{01} \dot\varphi_{00}
	,&
	\delta_{01} \lambda_{10}=&-\frac{i}{2}\epsilon_{01}\dot{\varphi}_{11}
	,\nonumber \\
	\delta_{01}  A_{11} =&-\frac{1}{2}\epsilon_{01}
	\dot\psi_{10}
	,&
	\delta_{01}  A_{00} =&-\frac{i}{2}\epsilon_{01}
	\dot{\lambda}_{01}
	. \label{NewVariTrans3}
\end{align}

Using the equations of motion \eqref{EoMA}, we get rid of $ V_{00}, V_{11}$ (instead of $A$'s) so that the coupling constant is absorbed into $A$'s and does not appear in $\mathcal{L}.$ 
We change the notations $ W_{00} = A_{00}, W_{11} = A_{11}$ as they will be the potentials of our model, then our Lagrangian reads
\begin{align}
	\mathcal{L} &= \frac{1}{2}(\dot{\varphi}_{00}^2+\dot{\varphi}_{11}^2)
	+i (\psi_{10} \dot{\psi}_{10} + \lambda_{10}  \dot{\lambda}_{10} + \psi_{01} \dot{\psi}_{01} + \lambda_{01}  \dot{\lambda}_{01}) 
	\nonumber \\
	&- 2W_{00}^2 - 2W_{11}^2 + 4(\psi_{10}\psi_{01} + \lambda_{10}\lambda_{01}) \partial_{00} W_{11} 
	+ 4i (\psi_{10}\lambda_{10} + \psi_{01} \lambda_{01}) \partial_{00} W_{00} 
	\label{1DLag}
\end{align}
and the constraints \eqref{ConstraintV} are given by
\begin{equation}
	\partial_{00} W_{00} = \partial_{11} W_{11}, \qquad 
	\partial_{00} W_{11} = \partial_{11} W_{00}. \label{ConstraintW}
\end{equation}
We here present an example of the potentials satisfying the constraints:
\begin{equation}
	W_{00} =  e^{\varphi_{00}} \cosh \varphi_{11}, 
	\qquad
	W_{11} = e^{\varphi_{00}} \sinh \varphi_{11}.
\end{equation}
The conserved charges corresponding to the transformations \eqref{NewVariTrans0}--\eqref{NewVariTrans3} are obtained from (3.47)--(3.50) of \cite{AiItoTa}:
\begin{align}
	H &= \frac{1}{2}(\dot{\varphi}_{00}^2 + \dot{\varphi}_{11}^2) + 2W_{00}^2 + 2W_{11}^2 \nn\\
	& \qquad- 4(\psi_{10}\psi_{01} + \lambda_{10}\lambda_{01})\partial_{00}W_{11} - 4i(\psi_{10}\lambda_{10} + \psi_{01}\lambda_{01})\partial_{00}W_{00} ,
	\nn \\
	Z &= 0,
	\nn \\
	Q_{10} &= \sqrt{2}(\dot{\varphi}_{00}\psi_{10} -i \dot{\varphi}_{11}\lambda_{01} + 2W_{00}\lambda_{10} + 2iW_{11}\psi_{01}) 
	\nn \\
	Q_{01} &= \sqrt{2} (\dot{\varphi}_{00}\psi_{01} -i \dot{\varphi}_{11}\lambda_{10} + 2W_{00}\lambda_{01} + 2iW_{11}\psi_{10}). \label{NCharges}
\end{align}
The charge $Z$ vanishes as the operator $Z$ does not generate any transformation, cf. \eqref{NewVariTrans0}. 

We now introduce the complex femionic variables:
\begin{equation}
	\xi := \psi_{10} + i \lambda_{10}, \qquad \eta := \psi_{01} + i \lambda_{01}.
\end{equation}
The Lagrangian \eqref{1DLag} becomes (up to total time derivative)
\begin{align}
	\mathcal{L} &= \frac{1}{2}(\dot{\varphi}_{00}^2+\dot{\varphi}_{11}^2) + i (\bar{\xi}\dot{\xi}  + \bar{\eta}\dot{\eta}) - 2W_{00}^2 - 2W_{11}^2
	\nonumber \\
	&+ 2(\bar{\eta} \xi + \bar{\xi} \eta) \partial_{00} W_{11} + 2(\bar{\xi}\xi + \bar{\eta}\eta)\partial_{00} W_{00}.
\end{align}
The equations of motion derived from the Lagrangian are given by
\begin{align}
	\ddot{\varphi_{00}} + 4W_{00} \partial_{00} W_{00} + 4 W_{11} \partial_{00} W_{11} -2(\bar{\eta} \xi + \bar{\xi} \eta) \partial_{00}^2 W_{11} - 2(\bar{\xi}\xi + \bar{\eta}\eta)\partial_{00}^2 W_{00} &= 0,
	\nonumber \\[4pt]
	\ddot{\varphi_{11}} + 4W_{00} \partial_{11} W_{00} + 4 W_{11} \partial_{11} W_{11} -2(\bar{\eta} \xi + \bar{\xi} \eta) \partial_{00}^2 W_{00} - 2(\bar{\xi}\xi + \bar{\eta}\eta)\partial_{00}^2 W_{11} &= 0,
	\nonumber \\[4pt]
	i\dot{\psi}_{10} + 2 \psi_{01} \partial_{00} W_{11} + 2i\lambda_{10} \partial_{00} W_{00} &= 0,
	\nonumber \\[4pt]
	i\dot{\lambda}_{10} + 2 \lambda_{01} \partial_{00} W_{11} - 2i\psi_{10} \partial_{00} W_{00} &=0, 
	\nonumber \\[4pt]
	i\dot{\psi}_{01} + 2 \psi_{10} \partial_{00} W_{11} + 2i\lambda_{01} \partial_{00} W_{00} &= 0,
	\nonumber \\[4pt]
	i\dot{\lambda}_{01} + 2 \lambda_{10} \partial_{00} W_{11} - 2i\psi_{01} \partial_{00} W_{00} &=0.
	\label{ELeq1}
\end{align}
In terms of the complex fermions, the conserved Noether charges $ Q_{10}$ and $ Q_{01}$  split into two parts which are conjugate each other, see \eqref{sc01}. 

\subsection{Hamiltonian formalism} \label{SEC:Hamilton}

When we switch from Lagrangian theory to Hamiltonian theory, we have to be careful about the order of $\Z2$-commutative variables and their derivatives, since the derivatives are also $\Z2$-commutative among themselves and have non-trivial relations with the $\Z2$-graded variables \cite{AKTqu}. 
We describe our conventions below. 

First, we define the conjugate momentum by
\begin{equation}
	p_q := \mathcal{L}\overleftarrow{\partial}_{q}, \quad 
	q \in \{ \varphi_{00}, \varphi_{11}, \xi, \bar{\xi}, \eta, \bar{\eta} \}
\end{equation}
Explicity
\begin{equation}
	p_{00} = \dot{\varphi}_{00}, \quad 
	p_{11} = \dot{\varphi}_{11}, \quad
	p_{\xi} = i\bar{\xi}, \quad p_{\eta} = i\bar{\eta}, \quad
	p_{\bar{\xi}} = p_{\bar{\eta}} = 0.  \label{ConjMom1D}
\end{equation}
We see that, as the standard supersymmetry, our model is a constrained system. 
We here employ the Dirac-Bergman method for constrained systems. 
The constraints are given by 
\begin{equation}
	\phi_{\xi} = p_{\xi} - i\bar{\xi}, \qquad 
	\phi_{\bar{\xi}} = p_{\bar{\xi}}, \qquad \phi_{\eta} = p_{\eta} - i\bar{\eta}, \qquad \phi_{\bar{\eta}} = p_{\bar{\eta}}.
	\label{constraintsH} 
\end{equation} 
The Hamiltonian and the total Hamiltonian involving the constraints are defined by
\begin{align}
	\mathcal{H} &= \sum_q p_q \dot{q} - \mathcal{L}
	\nonumber \\
	&= \frac{1}{2}(p_{00}^2+p_{11}^2) +2 W_{00}^2 + 2W_{11}^2 -2 (\bar{\eta }\xi + \bar{\xi} \eta) \partial_{00} W_{11} - \left([\bar{\xi} ,\xi] + [\bar{\eta} ,\eta]\right)\partial_{00} W_{00}
	\label{H1}
\end{align}
and 
\begin{equation}
	\mathcal{H}_T := \mathcal{H} + \alpha_{\xi} \phi_{\xi}  + \alpha_{\bar{\xi}} \phi_{\bar{\xi}} + \alpha_{\eta} \phi_{\eta}+ \alpha_{\bar{\eta}} \phi_{\bar{\eta}}
\end{equation}
where the Lagrange multiplier $\alpha_q$ has the degree same as $q.$ 
Then the Hamilton's equations of motion equivalent to the Euler-Lagrange equations \eqref{ELeq1} are given by
\begin{equation}
	\dot{q} = \overrightarrow{\partial}_{p_q} \mathcal{H}, \qquad 
	\dot{p_q} = -\mathcal{H} \overleftarrow{\partial}_q.
\end{equation}
The Hamiltonian \eqref{H1} is, of course, identical to the conserved Noether charge $H$ in \eqref{NCharges}. 
The supercharges in complex notations split into two parts:
\begin{align}
	Q_{10} &= \mathcal{Q}_{10} + \bar{\mathcal{Q}}_{10}, 
	\qquad 
   Q_{01} = \mathcal{Q}_{01} + \bar{\mathcal{Q}}_{01} \label{sc01}
\end{align}
with
\begin{align}
\mathcal{Q}_{10} &= \frac{1}{\sqrt{2}} \big( (p_{00}-2iW_{00}) \xi -(p_{11}-2i W_{11} )\eta \big),
\nn \\
\bar{\mathcal{Q}}_{10} &= \frac{1}{\sqrt{2}} \big( (p_{00}+2iW_{00}) \bar{\xi} +(p_{11}+2i W_{11} )\bar{\eta} \big),
\nn \\
\mathcal{Q}_{01} &= \frac{1}{\sqrt{2}} \big( (p_{00}-2iW_{00}) \eta - (p_{11}-2iW_{11}) \xi \big),
\nn \\
\bar{\mathcal{Q}}_{01} &= \frac{1}{\sqrt{2}} \big( (p_{00} + 2iW_{00})\bar{\eta} + (p_{11}+2iW_{11})\bar{\xi} \big).
\label{N2supercharge}
\end{align}

Now we introduce the $\Z2$-version of the Poisson bracket
\begin{equation}
	\PB{A}{B} := A \hat{\Gamma} B - (-1)^{\vec{a}\cdot\vec{b}} B \hat{\Gamma} A, \qquad
	\hat{\Gamma} := \sum_q \overleftarrow{\partial}_{q} \overrightarrow{\partial}_{p_q}, \qquad 
	\vec{a} := \deg{A}.
\end{equation}
It is straightforward to verify that the Poisson bracket satisfy the following relations:
\begin{align}
	\PB{A}{B} &= -\ph{a}{b} \PB{B}{A},
	\nn \\
	\PB{A}{BC} &= \PB{A}{B}C + \ph{a}{b} B\PB{A}{C},
	\nn\\
	\PB{A}{\PB{B}{C}} &= \PB{\PB{A}{B}}{C} + \ph{a}{b} \PB{B}{\PB{A}{C}}.
	\label{BracketProp}
\end{align}
The constraints \eqref{constraintsH} are the second class as there exist non-vanishing Poisson brackets:
\begin{equation}
	\PB{\phi_{\xi}}{\phi_{\bar{\xi}}} = \PB{\phi_{\eta}}{\phi_{\bar{\eta}}} = -i.
\end{equation}
The time evolution of the constraints determined by the equation $ \dot{\phi}_{q} = \PB{\phi_{q}}{\mathcal{H}_T}$ is summarized as
\begin{align}
	(\dot{\phi}_{\xi}, \dot{\phi}_{\bar{\xi}}, \dot{\phi}_{\eta}, \dot{\phi}_{\bar{\eta}}) 
	&= (\PB{\phi_{\xi}}{\mathcal{H}}, \PB{\phi_{\bar{\xi}}}{\mathcal{H}}, \PB{\phi_{\eta}}{\mathcal{H}}, \PB{\phi_{\bar{\eta}}}{\mathcal{H}}) 
	\nonumber \\
	&+ (\alpha_{\xi}, \alpha_{\bar{\xi}}, \alpha_{\eta}, \alpha_{\bar{\eta}}) \Delta = 0
\end{align}
where
\begin{align}
	\Delta &:= 
	\left(
	\begin{array}{rr|rr}
		-\PB{\phi_{\xi}}{\phi_{\xi}} & -\PB{\phi_{\bar{\xi}}}{\phi_{\xi}} & \PB{\phi_{\eta}}{\phi_{\xi}} & \PB{\phi_{\bar{\eta}}}{\phi_{\xi}}
		\\[4pt]
		-\PB{\phi_{\xi}}{\phi_{\bar{\xi}}} & -\PB{\phi_{\bar{\xi}}}{\phi_{\bar{\xi}}} & \PB{\phi_{\eta}}{\phi_{\bar{\xi}}} & \PB{\phi_{\bar{\eta}}}{\phi_{\bar{\xi}}}
		\\[4pt] \hline
		\PB{\phi_{\xi}}{\phi_{\eta}} & \PB{\phi_{\bar{\xi}}}{\phi_{\eta}} & -\PB{\phi_{\eta}}{\phi_{\eta}} & -\PB{\phi_{\bar{\eta}}}{\phi_{\eta}}
		\\[4pt]
		\PB{\phi_{\xi}}{\phi_{\bar{\eta}}} & \PB{\phi_{\bar{\xi}}}{\phi_{\bar{\eta}}} & -\PB{\phi_{\eta}}{\phi_{\bar{\eta}}} & -\PB{\phi_{\bar{\eta}}}{\phi_{\bar{\eta}}}	  	
	\end{array}	
	\right)
	\nonumber \\[4pt]
	&= i
	\left(
	\begin{array}{cc}
		\sigma_1 & 0 \\ 0 & \sigma_1
	\end{array}	
	\right).
\end{align}
This relations determine the Lagrange multiplier
\begin{align}
	(\alpha_{\xi}, \alpha_{\bar{\xi}}, \alpha_{\eta}, \alpha_{\bar{\eta}}) 
	&= -(\PB{\phi_{\xi}}{\mathcal{H}}, \PB{\phi_{\bar{\xi}}}{\mathcal{H}}, \PB{\phi_{\eta}}{\mathcal{H}}, \PB{\phi_{\bar{\eta}}}{\mathcal{H}}) \Delta^{-1}
	\nonumber \\
	&=i (\PB{\phi_{\bar{\xi}}}{\mathcal{H}}, \PB{\phi_{\xi}}{\mathcal{H}}, \PB{\phi_{\bar{\eta}}}{\mathcal{H}}, \PB{\phi_{\eta}}{\mathcal{H}}).
\end{align}
More explicitly, we have the expressions:
\begin{align}
	\alpha_{\xi} &= -2i\eta\, \partial_{00} W_{11} -2i \xi\, \partial_{00} W_{00},
	\nonumber \\
	\alpha_{\bar{\xi}} &= -2i\bar{\eta}\, \partial_{00} W_{11} +2i \bar{\xi}\, \partial_{00} W_{00} = \overline{\alpha}_{\xi},
	\nonumber \\
	\alpha_{\eta} &= -2i \xi\, \partial_{00} W_{11} -2i \eta\, \partial_{00} W_{00},
	\nonumber \\
	\alpha_{\bar{\eta}} &= -2i \bar{\xi}\, \partial_{00} W_{11} +2i \bar{\eta}\, \partial_{00} W_{00} = \overline{\alpha}_{\eta}.
\end{align}

With these data one may defined a $\Z2$-version of the Dirac bracket by
\begin{align}
	\DB{A}{B} &:= \PB{A}{B} + \sum_{q,q'} \PB{A}{\phi_q}\, \Delta^{-1}_{qq'} \, \PB{\phi_{q'}}{B}
	\nonumber \\
	&= \PB{A}{B} -i \PB{A}{\phi_{\xi}} \PB{\phi_{\bar{\xi}}}{B} -i \PB{A}{\phi_{\bar{\xi}}} \PB{\phi_{\xi}}{B} 
	\nonumber \\
	&-i \PB{A}{\phi_{\eta}} \PB{\phi_{\bar{\eta}}}{B} -i \PB{A}{\phi_{\bar{\eta}}} \PB{\phi_{\eta}}{B}.
	\label{DBDef}
\end{align}
It is not difficult to verify that the Dirac bracket satisfies the same relations \eqref{BracketProp} as the $\Z2$-Poisson bracket.

One may easily find that the non-vanishing Dirac brackets for the canonical variables are the followings
\begin{equation}
	\DB{\varphi_{00}}{p_{00}} = \DB{\varphi_{11}}{p_{11}} = \DB{\xi}{p_{\xi}} = \DB{\eta}{p_{\eta}} = 1.
\end{equation}
Using \eqref{ConjMom1D}, the Dirac brackets for the fermionic variables are converted into the form:
\begin{equation}
	\DB{\xi}{\bar{\xi}} = \DB{\eta}{\bar{\eta}} = -i.
\end{equation}

We introduce the quantity of  $\Z2$-degree $(1,1)$:
\begin{align}
	 \mathcal{Z} =
	-p_{00}p_{11} -4W_{00}W_{11}
	+2\partial_{00} W_{00}(\bar{\xi} \eta +\bar{\eta} \xi)
	+\partial_{00}  W_{11}([\bar{\xi}, \xi] +[\bar{\eta}, \eta]).
	\label{Z-classical}
\end{align}
Then one may verify that $ \mathcal{H}, \mathcal{Q}_{a}, \bar{\mathcal{Q}}_{a}, \mathcal{Z}$ close in the $\mathcal{N}=2$ extended $\Z2$-supersymmetry algebra whose non-vanishing Dirac brackets are given by
\begin{align}
		\DB{\mathcal{Q}_{10}}{\bar{\mathcal{Q}}_{10}} &= \DB{\mathcal{Q}_{01}}{\bar{\mathcal{Q}}_{01}} = -i\mathcal{H},
		\nn \\
		\DB{\bar{\mathcal{Q}}_{10}}{\mathcal{Q}_{01}} &= -\DB{\mathcal{Q}_{10}}{\bar{\mathcal{Q}}_{01}} = i\mathcal{Z}.
\end{align}
The combined $\mathcal{N} = 1 $ supercharges \eqref{sc01} satisfy the $\mathcal{N} = 1$ $\Z2$-supersymmetry algebra with vanishing $Z$:
\begin{equation}
	\DB{Q_{10}}{Q_{10}} = \DB{Q_{01}}{Q_{01}} = -2i\mathcal{H}, \qquad \DB{Q_{10}}{Q_{01}} = 0.
\end{equation}

\section{$\mathcal{N}=2$ $\Z2$-supersymmetric quantum mechanics} \label{SEC:QM}
\setcounter{equation}{0}


We quantize the system discussed in \S \ref{SEC:Hamilton} which means that the Dirac bracket is replaced with the $\Z2$-Lie bracket ($\hbar =1$):
\begin{equation}
	\DB{A}{B} \ \to \ \frac{1}{i}\llbracket A, B \rrbracket
\end{equation}
This gives the following non-vanishing (anti)commutators
\begin{equation}
	[\varphi_{00}, p_{00}] = [\varphi_{11}, p_{11}] = i, \qquad
	\{\xi, \xi^\dag \} = \{ \eta, \eta^\dag \} = 1 \label{canonical_rel}
\end{equation}
and all the followings vanish
\begin{alignat}{4}
	& \{ \xi, \xi\}, & \qquad & \{ \xi^\dag, \xi^\dag \}, & \qquad &\{ \eta, \eta\}, 
	& \qquad
	& \{ \eta^\dag, \eta^\dag\},
	\nonumber
	\\
	&[\xi,\eta], & & [\xi, \eta^\dag], & & [\xi^\dag, \eta], & & [\xi^\dag,\eta^\dag],
	\nonumber \\
	&\{c_{11}, \xi \}, & & \{c_{11}, \eta \}, & &\{c_{11}, \xi^\dag \}, & & \{c_{11}, \eta^\dag \}, \quad c_{11} = \varphi_{11}, p_{11}
	\label{canonical_rel2}
\end{alignat}
where and in what follows we use ``dagger", instead of ``bar", for the hermitian conjugation of the quantum operators. 

By using the real representation of the Clifford algebra $Cl(4,2)$ \cite{SO1,SO2,CRTop}, 
the relations \eqref{canonical_rel} and \eqref{canonical_rel2} are realized by matrix differential operators. 
In this realization, the $\Z2$-grading is carried by the matrices which means that  
if there are non-zero entries in one of the following blocks, the matrix has the indicated $\Z2$-degree:
\begin{equation}
	\begin{pmatrix}
		(0,0) & (1,1) & (1,0) & (0,1) \\
		(1,1) & (0,0) & (0,1) & (1,0) \\
		(1,0) & (0,1) & (0,0) & (1,1) \\
		(0,1) & (1,0) & (1,1) & (0,0)
	\end{pmatrix}.
\end{equation}
The Clifford algebra $Cl(4,2)$ is generated by $ \gamma_i, i= 1, 2, \dots 6$ which subject to the relations
\begin{align}
	\{\gamma_i, \gamma_j\}=2\eta_{ij}, \quad\eta=\mathrm{diag}(1,1,1,1,-1,-1).
\end{align}
We introduce the anticommuting matrices $ X, Y, A $ and the identity matrix $I$:
\begin{equation}
	I:=\begin{pmatrix}
		1 & 0 \\ 0 & 1
	\end{pmatrix},
	\quad
	X:= \begin{pmatrix}
		1 &  0 \\ 0 & -1
	\end{pmatrix},
	\quad
	Y := \begin{pmatrix}
		0 & 1 \\ 1 & 0
	\end{pmatrix},
	\quad
	A := \begin{pmatrix}
		0 & 1 \\ -1 & 0
	\end{pmatrix}\label{alpha_rep}
\end{equation}
then the real irrep of $Cl(4,2)$ is given by 
\begin{alignat}{3}
	\gamma_1 &= XII, \ (0,0), & \qquad \gamma_2&=YII, \ (1,0), & \qquad \gamma_3 &= AAI, \ (0,1),
	\nn \\
	\gamma_4 &=AYA, \ (0,1), & \gamma_5 &= AXI, \ (1,0), & \gamma_6 &= AYX, \ (0,1)
\end{alignat}
where a word consisting of these matrices is understand as the tensor product, e.g. 
$ XYA = X \otimes Y \otimes A$ and the $\Z2$-degree of $\gamma_i$ is also indicated. 
With this eight-dimensional irrep, the $\mathbb{Z}_2^2$-graeded quantum oparators are realized as:
\begin{alignat}{2}
	\varphi_{00}=& x_0 I_8,&\quad p_{00}=&-i\partial_{x_0} I_8,\\
	\varphi_{11}=& x_1 \Gamma,&\quad p_{11}=&-i\partial_{x_1}\Gamma,\\
	\xi=&\frac{i}{2}(\gamma_1\gamma_5+i\gamma_3\gamma_4\gamma_5),&\quad \xi^\dag=&-\frac{i}{2}(\gamma_1\gamma_5-i\gamma_3\gamma_4\gamma_5) \\
	\eta=&\frac{1}{2}(\gamma_3+i\gamma_4),&\quad \eta^\dag=&\frac{1}{2}(\gamma_3-i\gamma_4), \label{m-reps}
\end{alignat}
where $I_8=III$, $\Gamma=-\gamma_3\gamma_4\gamma_5\gamma_6$, and $x_0,\, x_1 \in \mathbb{R}.$ 
The degree $(1,1)$ function $ W_{11} $ is also realized by the matrix $\Gamma$ and the constraints  \eqref{ConstraintW} read as follows:
\begin{equation}
	W_{11} = \tilde W_{00}(x_0,x_1) \Gamma,
	\qquad
	\partial_{x_0} W_{00}=\partial_{x_1}\tilde W_{00}
	,\qquad
	\partial_{x_1} W_{00}=\partial_{x_0}\tilde W_{00}
	\label{two_conditions}	
\end{equation}
where $\tilde{W}_{00}$ is a degree $(0,0)$ function. 
Therefore, we get two-dimensional or two-particle (same mass) quantum mechanical system in this realization.

The quantized $\mathcal{N}=2$ supercharges \eqref{N2supercharge} are given by
\begin{alignat}{2}
	\mathcal{Q}_{10} &= a\xi -b\Gamma\eta
	,\qquad&
	\mathcal{Q}^\dag_{10} &= a^\dag \xi^\dag +b^\dag \Gamma\eta^\dag,
	\\
	\mathcal{Q}_{01} &= a \eta - b\Gamma\xi
	,\qquad&
	\mathcal{Q}^\dag_{01} &= a^\dag \eta^\dag +b^\dag \Gamma \xi^\dag
	,
\end{alignat}
where 
$a:=\dfrac{1}{\sqrt{2}}(-i\partial_{x_0}-2iW_{00})$, $b:=\dfrac{1}{\sqrt{2}}(-i\partial_{x_1}-2i\tilde{W}_{00})$.

We introduce the new operators 
\begin{alignat}{2}
	A:=&\frac{1}{\sqrt{2}}(a+b), \quad &B:=&\frac{1}{\sqrt{2}}(a-b),\label{op_AB}
\end{alignat}
and the unitary matrix which diagonalize the Hamiltonian \eqref{H1}
\begin{equation}
	U = \left(
	\begin{array}{cc:cc|cc:cc}
		1& 0& & & & & & \\
		0& 1& & & & & & \\ \hdashline
		& & 0& 1& & & & \\
		& & 1& 0& & & & \\ \hline
		& & & & \frac{i}{\sqrt{2}} & -\frac{i}{\sqrt{2}}& & \\
		& & & & \frac{1}{\sqrt{2}} & \frac{1}{\sqrt{2}} & & \\ \hdashline
		& & & & & & -\frac{i}{\sqrt{2}}&\frac{i}{\sqrt{2}} \\ 
		& & & & & & \frac{1}{\sqrt{2}} & \frac{1}{\sqrt{2}}
	\end{array}
	\right).
\end{equation}
Then, we have the $\mathcal{N} = 2$  $\Z2$-SQM
\begin{align}
	\tilde{\mathcal{H}} := U^{\dagger} \mathcal{H} U = \mathrm{diag} (H_1, H_2, H_1, H_2, H_3, H_4, H_3, H_4)
	\label{H_tilde}
\end{align}
where
\begin{alignat}{2}
	H_1=&A A^\dag + B^\dag B
	,\qquad&
	H_2=&A^\dag A + B B^\dag
	,\nn\\
	H_3=&A A^\dag + B B^\dag
	,\qquad&
	H_4=&A^\dag A + B^\dag B
	\label{comp_H}
\end{alignat}
with the supercharges
\begin{align}
	\tilde{ \mathcal{Q}}_{10}
	:=& U^{\dagger} \mathcal{Q}_{10} U
	= \left(
	\begin{array}{cc:cc|cc:cc}
		& & & & 0& A& & \\
		& & & & 0& iB& & \\\hdashline
		& & & & & & 0& -iA\\
		& & & & & & 0& B\\\hline
		B& iA& & & & & & \\
		0& 0& & & & & & \\\hdashline
		& & -iB& A& & & & \\ 
		& & 0& 0& & & & 
	\end{array}
	\right),
	\\[3pt]
	\tilde{\mathcal{Q}}_{01}
	:=& U^{\dagger} \mathcal{Q}_{01} U
	= \left(
	\begin{array}{cc:cc|cc:cc}
		& & & & & & 0& iA\\
		& & & & & & 0& B\\\hdashline
		& & & & 0& -A& & \\
		& & & & 0& iB& & \\\hline
		& & -B& iA& & & & \\
		& & 0& 0& & & & \\\hdashline
		iB& A& & & & & & \\
		0& 0& & & & & & 
	\end{array}
	\right) 
\end{align} 
and their hermitian conjugation. 
Furthermore, we have the non-vanishing degree $(1,1)$ operator \eqref{Z-classical}
\begin{align}
	\tilde{\mathcal{Z}} :=&
	U^{\dagger} \mathcal{Z} U 
	=\left(
	\begin{array}{cc:cc|cc:cc}
		& & Z_1& & & & & \\
		& & & Z_2& & & & \\\hdashline
		Z_1^\dag& & & & & & & \\
		& Z_2^\dag& & & & & & \\\hline
		& & & & & & Z_3& \\
		& & & & & & & Z_4\\\hdashline
		& & & & Z_3^\dag& & & \\
		& & & & & Z_4^\dag& & 
	\end{array}
	\right)
	\label{Z_tilde}
\end{align}
where
\begin{alignat}{2}
	Z_1 = &-A A^\dag + B^\dag B = Z_1^\dag 
	,\quad&
	Z_2 = &-A^\dag A + B B^\dag = Z_2^\dag
	,\nn\\
	Z_3 = & i( A A^\dag - B B^\dag ) = -Z_3^\dag
	,\quad&
	Z_4 = & -i( A^\dag A - B^\dag B )= -Z_4^\dag.
\end{alignat}
The products of $ A, A^{\dagger}$ and $ B, B^{\dagger}$ are given by
\begin{align}
	A^\dag A &= -\frac{1}{4}(
	\partial_{x_0}+\partial_{x_1}
	)^2
	+W_{00}^2+\tilde W_{00}^2
	-\partial_{x_0}W_{00}-\partial_{x_0}\tilde W_{00}+2W_{00}\tilde W_{00},
	\nn \\
	A A^\dag &= A^\dag A
	+ 2\partial_{x_0}W_{00}+2\partial_{x_0}\tilde W_{00} 
	\label{ProductAA}
\end{align}
and
\begin{align}
	B^\dag B &= -\frac{1}{4}(
	\partial_{x_0}-\partial_{x_1}
	)^2
	+W_{00}^2+\tilde W_{00}^2
	-\partial_{x_0}W_{00}+\partial_{x_0}\tilde W_{00}-2W_{00}\tilde W_{00},
	\nn \\
	B B^\dag &= B^\dag B + 2\partial_{x_0}W_{00}-2\partial_{x_0}\tilde W_{00}
	\label{ProductBB}
\end{align}
where we used \eqref{two_conditions} to have these formulae. 
The relations \eqref{two_conditions} are also used to see that the non-vanishing commutation relations among $A^\dag,A,B^\dag ,B$ are following:
\begin{alignat}{2}
	[A, A^\dag]=&2\partial_{x_0}W_{00}+2\partial_{x_0}\tilde W_{00}
	,\quad&
	[B, B^\dag]=&2\partial_{x_0}W_{00}-2\partial_{x_0}\tilde W_{00}
	.
\end{alignat}

It is not difficult to verify that $ \tilde{\mathcal{H}}, \tilde{\mathcal{Q}}_{a}, \tilde{\mathcal{Q}}_{a}^{\dagger}$ and $\tilde{\mathbb{Z}} $ forms the  $\mathcal{N}=2$ $\Z2$-supersymmetry algebra whose non-vanishing relations are given by 
\begin{equation}
	\{\tilde{\mathcal{Q}}_{10},\tilde{\mathcal{Q}}_{10}^\dag\}= \{\tilde{\mathcal{Q}}_{01},\tilde{\mathcal{Q}}_{01}^\dag\}=\tilde{\mathcal{H}},\qquad [\tilde{\mathcal{Q}}_{10},\tilde{\mathcal{Q}}_{01}^\dag]=-[\tilde{\mathcal{Q}}_{10}^\dag,\tilde{\mathcal{Q}}_{01}]=\tilde{\mathcal{Z}}
	.\label{N2alg}
\end{equation} 
It is also immediate that the combined $\mathcal{N} = 1 $ supercharges \eqref{sc01} satisfy the $\mathcal{N} = 1$ $\Z2$-supersymmetry algebra with vanishing $Z$:
\begin{equation}
   \{Q_{10},Q_{10}\}=\{Q_{01},Q_{01}\}=2\mathcal{H},
	\qquad 
	[Q_{10},Q_{01}] =0. 
\end{equation}
One may also see from \eqref{H_tilde} and \eqref{Z_tilde} that $ \tilde{\mathcal{Z}}^2 \neq \tilde{\mathcal{H}}^2. $ 
This is the sharp contrast to the $\Z2$-SQMs discussed in the literature \cite{BruDup,AAD,AKTqu} where one always observe that $Z^2 = H^2.$ 
The relation $ \tilde{\mathcal{Z}}^2 \neq \tilde{\mathcal{H}}^2 $  implies that our $\Z2$-SQM is a quantum mechanical realization of a eight dimensional irrep of $\mathcal{N} = 2 $ $\Z2$-supersymmetry algebra. 
In \cite{AiDoi}, it is shown that irreps of the $\mathcal{N}=2$ $\Z2$-supersymmetry algebra are four-dimensional if $ \tilde{\mathcal{Z}}^2 = \tilde{\mathcal{H}}^2, $ but eight-dimensional otherwise. 
Our $\Z2$-SQM is the first example of the physical realization of eight-dimensional irrep of the $\Z2$-supersymmetry algebra.

The formulae \eqref{ProductAA} and \eqref{ProductBB} suggest the introduction of the light cone coordinates
\begin{equation}
	 x_+:= x_0+x_1
	,\qquad
	x_-:=x_0-x_1.
\end{equation}
The constraints in \eqref{two_conditions} become
\begin{align}
	\partial_+ W_{00}(x_+,x_-)=&\partial_+ \tilde W_{00}(x_+,x_-)
	,\\
	\partial_- W_{00}(x_+,x_-)=&-\partial_- \tilde W_{00}(x_+,x_-)
\end{align}
and these differential equations may be solved to give the separation of left and right movers 
\begin{align}
	W_{00}(x_+,x_-)=&\frac{1}{2}\left(W^\prime_+(x_+)+W^\prime_-(x_-)\right)
	,\\
	\tilde W_{00}(x_+,x_-)=&\frac{1}{2}\left(W'_+(x_+)-W'_-(x_-)\right)
\end{align}
where the prime stands for the derivative. 
The operators \eqref{op_AB} in the light cone coordinates yield the standard ones in the SQM:
\begin{align}
	A=-i\partial_+ -iW'_+
	,\qquad
	B=-i\partial_- -iW'_-
\end{align}
which give the followings
\begin{alignat}{2}
	A^{\dag} A &= -\partial_+^2 +(W'_+)^2 -W''_+, & \qquad
	A A^\dag &= -\partial_+^2 +(W'_+)^2 +W''_+ ,
	\\ \nn
	B^\dag B &= - \partial_-^2 +(W'_-)^2 -W''_-, & 
	B B^\dag &= - \partial_-^2 +(W'_-)^2 +W''_-.
\end{alignat}

The Hilbert space of our $\Z2$-SQM  is $ \mathfrak{H} = L^2(\mathbb{R}) \otimes \mathbb{C}^8$ and the space is also $\Z2$-graded:
\begin{equation}
	\mathfrak{H} = \mathfrak{H}_{(0,0)} \oplus \mathfrak{H}_{(1,1)} \oplus \mathfrak{H}_{(1,0)}\oplus \mathfrak{H}_{(0,1)}. 
\end{equation}
The algebra \eqref{N2alg} implies that the Hamilotonian $\tilde{\mathcal{H}}$ \eqref{H_tilde} is positive semi-definite. This is also seen from the component Hamiltonian  $H_k $ \eqref{comp_H} all of which are  also positive semi-definite. 
The zero energy ground state $\Psi_0 $ of $ \tilde{\mathcal{H}} $ is determined by  
\begin{equation}
	\tilde{ \mathcal{Q}}_a \Psi_0 = \tilde{ \mathcal{Q}}_a^\dag \Psi_0 = 0.  
\end{equation}
This is equivalent to finding the zero energy states of the component Hamiltonian $H_k \psi_0^{(k)} = 0$. More explicitly, $\psi_0^{(k)}$ are solutions of the equations
\begin{align}
	A^\dag \psi_0^{(1)} &= B\psi_0^{(1)} = 0, & 
	A \psi_0^{(2)} &= B^\dag \psi_0^{(2)} = 0,
	\nn\\
	A^\dag \psi_0^{(3)} &= B^\dag \psi_0^{(3)} = 0, & 
    A \psi_0^{(4)} &= B \psi_0^{(4)} = 0.	
\end{align}
It is easy to solve these equations:
\begin{align}
	\psi_0^{(1)}=& \exp\left(W_+\right) \exp\left(-W_-\right)
	,\quad &
	\psi_0^{(2)}=& \exp\left(-W_+\right) \exp\left(W_-\right)
	,\nonumber \\
	\psi_0^{(3)}=& \exp\left(W_+\right) \exp\left(W_-\right)
	,\quad &
	\psi_0^{(4)}=& \exp\left(-W_+\right) \exp\left(-W_-\right).
\end{align}
It is also easy to see that only one of them is normalizable. 
For instance, if $ \psi_0^{(1)}$ is normalizable, all other functions are not normalizable. 
Therefore, the possible ground state is one of the followings $(c \in \mathbb{C}$ is a constant)
\begin{align}
	& (\psi_0^{(1)}, 0, c \,\psi_0^{(1)}, 0, 0, 0, 0, 0) \ \in \ \mathfrak{H}_{(0,0)} \oplus \mathfrak{H}_{(1,1)}
    \nonumber\\
   & ( 0, \psi_0^{(2)}, 0, c \,\psi_0^{(2)},  0, 0, 0, 0)\  \in \ \mathfrak{H}_{(0,0)} \oplus \mathfrak{H}_{(1,1)}
    \nonumber \\
    &  ( 0, 0, 0, 0, \psi_0^{(3)}, 0, c \,\psi_0^{(3)},  0)\  \in \ \mathfrak{H}_{(1,0)} \oplus \mathfrak{H}_{(0,1)}	
    \nonumber \\
    &  ( 0, 0, 0, 0, 0,\psi_0^{(4)}, 0, c\,\psi_0^{(4)}) \ \in \ \mathfrak{H}_{(1,0)} \oplus   \mathfrak{H}_{(0,1)}.	
\end{align}
Therefore, the ground state is either non-existent or two-fold degenerate and belongs to  $ \mathfrak{H}_{(0,0)} \oplus \mathfrak{H}_{(1,1)} $ or $ \mathfrak{H}_{(1,0)} \oplus \mathfrak{H}_{(0,1)}. $

%
\section{Concluding remarks} \label{SEC:CR}

In order to investigate a quantum theory relating to the  the $\mathcal{N} =1$ $\Z2$-supersymmetric Lagrangian \eqref{Lint2D}, we studied the $\Z2$-SQM obtained from the Lagrangian by dimensional reduction. 
The dimensional reduction increases the supersymmetry from $\mathcal{N} = 1 $ to $ \mathcal{N} = 2 $ and we employed the $\Z2$-extended Dirac-Bargmann method to quantize the system. 
The $\Z2$-SQM obtained is a two-dimensional or two-particle quantum system in which the right and left movers are separated. 
It is also a quantum mechanical realization of the eight-dimensional irrep of $\mathcal{N} = 2$ $\Z2$-supersymmetry algebra discussed in \cite{AiDoi}. 
Moreover, it is the first $\Z2$-SQM with $ \tilde{\mathcal{Z}}^2 \neq \tilde{\mathcal{H}}^2. $ 

A conformal extension of the present $\Z2$-SQM will not be difficult, since there is a large freedom of choice of the super potentials. An appropriate choice of them will give a quantum mechanical realization of the $\mathcal{N} = 2$ $\Z2$-superconformal algebra whose representation theory has not been studied in detail. 
We comment that irreps of the $\mathcal{N}=1$ $\Z2$-superconformal algebra ($\Z2$-$osp(1|2)$) are studied in detail in \cite{NAKA}. 
The $\mathcal{N} = 2 $ $\Z2$-superconformal mechanics and the related representation theory are interesting future work. 

It is also interesting to investigate some special choices of the superpotential $W_{\pm}(x_{\pm}),$ e.g., harmonic oscillator, since we may have a larger symmetry. 
In \cite{BecHus}, it is shown that the largest spectrum generating algebra of the supersymmetric harmonic oscillator is the semidirect some of $osp(2|2)$ and 1D Heisenberg superalgebra. 
However, one may easily verify that the operators in the article also close in a $\Z2$-graded Lie superalgebra. If we consider the $\Z2$-supersymmetric harmonic oscillator, then the largest spectrum generating algebra will be higher graded than the $\Z2$-grading. 


\section*{Acknowledgments} 

N. A. is supported by JSPS KAKENHI Grant Number JP23K03217. 

%
%

\end{document}